\documentclass[12pt,preprint,number,sort&compress,lontitle,times]{elsarticle}
\usepackage{natbib}
\usepackage{graphicx}
\usepackage{epsfig}
\usepackage{lineno}
\usepackage{multirow}
\usepackage{fourier}
\usepackage{color}
\usepackage{appendix}
\usepackage[english]{babel}
\usepackage[]{geometry}
\usepackage{hyperref}
\newcommand{\dif}{\mathrm{d}}

\begin{document}

\begin{frontmatter}

\title{The Small Contribution of Molecular Bremsstrahlung \\
Radiation to the Air-Fluorescence Yield \\ 
of Cosmic Ray Shower Particles}
\author[lpnhe]{Imen Al Samarai}
\ead{ialsamar@lpnhe.in2p3.fr}
\author[ipn]{Olivier Deligny}
\ead{deligny@ipno.in2p3.fr}
\author[madrid]{Jaime Rosado}
\ead{jaime\_ros@fis.ucm.es}
\address[lpnhe]{Laboratoire de Physique Nucl\'eaire et des Hautes \'Energies, \\
CNRS/IN2P3 \& Universit\'{e}s Pierre et Marie Curie and Paris Diderot, 75252 Paris Cedex, France}
\address[ipn]{Institut de Physique Nucl\'eaire,\\ CNRS-IN2P3, Univ. Paris-Sud, Universit\'{e} Paris-Saclay, 91406 Orsay Cedex, France}
\address[madrid]{Departamento de F\'{i}sica At\'{o}mica, Molecular y Nuclear, Facultad de Ciencias F\'{i}sicas,\\ Universidad Complutense de Madrid, E-28040 Madrid, Spain}

\begin{abstract}
A small contribution of molecular Bremsstrahlung radiation to the air-fluorescence yield in the UV range is estimated based on an approach previously developed in the framework of the radio-detection of showers in the gigahertz frequency range. First, this approach is shown to provide an estimate of the main contribution of the fluorescence yield due to the de-excitation of the C~$^3\Pi_{\mathrm{u}}$ electronic level of nitrogen molecules to the B~$^3\Pi_{\mathrm{g}}$ one amounting to $Y_{[337]}=(6.05\pm 1.50)~$MeV$^{-1}$ at 800~hPa pressure and 293~K temperature conditions, which compares well to previous dedicated works and to experimental results. Then, under the same pressure and temperature conditions, the fluorescence yield induced by molecular Bremsstrahlung radiation is found to be $Y_{[330-400]}^{\mathrm{MBR}}=0.10~$MeV$^{-1}$ in the wavelength range of interest for the air-fluorescence detectors used to detect extensive air showers induced in the atmosphere by ultra-high energy cosmic rays. This means that out of $\simeq 175~$photons with wavelength between 330 and 400~nm detected by  fluorescence detectors, one of them has been produced by molecular Bremsstrahlung radiation. Although small, this contribution is not negligible in regards to the total budget of systematic uncertainties when considering the absolute energy scale of fluorescence detectors. 
\end{abstract}
\end{frontmatter}

\section{Introduction}

Ultra-high energy cosmic rays can be efficiently observed by collecting the isotropic atmospheric fluorescence light emitted during dark nights by the extensive air showers that they induce. This detection technique, pioneered by the Fly's Eye experiment~\cite{FlysEye}, was used by the HiRes experiment~\cite{HiRes} and is currently used at the Pierre Auger Observatory~\cite{Auger} and the Telescope Array~\cite{TA}. Fluorescence telescopes allow an accurate observation of the longitudinal profile of the showers. Based on the assumption that the fluorescence intensity is proportional to the electromagnetic energy deposited by the shower in the atmosphere, the fluorescence detection technique provides the opportunity of performing shower calorimetry by mapping the ionisation content along the shower tracks. It is, as of today, the most direct way to measure the primary energy on a nearly model-independent basis.

The various processes leading to the air-fluorescence emission have been well investigated theoretically in the recent years~\cite{Blanco2005,Arqueros2009,Rosado2014}. The atmospheric fluorescence produced in an extensive air shower is now known to be due mainly to the de-excitation of nitrogen molecules previously excited by the numerous low-energy ionisation electrons left in the atmosphere after the passage of the electrons/positrons of the shower front. Low-energy ionisation electrons can in turn produce their own emission through the molecular Brem\-sstrah\-lung process, as a consequence of their collisions with nitrogen or oxygen targets. The corresponding radiation is expected to occur over a wide range of frequencies; and the purpose of this work is to quantify the contribution of the molecular Bremsstrahlung radiation to the air-fluorescence light emitted in the UV wavelength range. 

The general approach follows from previous works dedicated to the emission in the gigahertz frequency range~\cite{AlSamarai2015,AlSamarai2016}, applied here to the UV wavelength range. To start with, a comprehensive treatment of the time evolution of the distribution function describing the low-energy ionisation electrons produced after the passage of high-energy electrons in air is presented in section~\ref{sec:production}. From the knowledge of this distribution function, the fluorescence yield expected from the de-excitation of nitrogen molecules is estimated in section~\ref{sec:fy} and is shown to compare favorably to the various measurements under 800~hPa pressure and 293~K temperature conditions. Then, with the same pressure and temperature conditions, the application of the formalism to the fluorescence yield expected from molecular Bremsstrahlung radiation is presented in section~\ref{sec:fy_mbr}. Finally, conclusions are given in section~\ref{sec:conclusions}.


\section{Production and Interactions of Low-Energy Ionisation Electrons}
\label{sec:production}

Throughout this study, the case of primary electrons with typical energy of a few tens of mega-electron volts is considered. Upon the passage of such high-energy electrons in air, there is roughly 2 MeV~g$^{-1}$~cm$^{2}$ of deposited energy per grammage unit mainly through the ionisation process. Hence, for one primary electron with energy $T_\mathrm{p}$ propagating over an infinitesimal distance $\dif x$ in air, and for a mass density and average molar mass of dry air of target molecules $\rho$ and $A$, the average number of ionisation electrons per unit length and per kinetic energy band can be estimated as
\begin{equation}
\label{eqn:d2N_dxdT}
\frac{\dif^2n_{\mathrm{e,i}}(T_\mathrm{p},T)}{\dif x~\dif T}=\frac{\rho \mathcal{N}}{A}\sigma_{\mathrm{ion}}(T_\mathrm{p})f_0(T_\mathrm{p},T),
\end{equation}
with $\mathcal{N}$ the Avogadro number. The ionisation cross section $\sigma_{\mathrm{ion}}(T_\mathrm{p})$ is considered identical for nitrogen or oxygen targets in this study. $f_0(T)$ stands for the normalised distribution in kinetic energy of the ionisation electrons at their time of creation, $f_0(T)\equiv f(T,t=0)$. This distribution has been experimentally determined and accurately parameterised for primary high-energy electrons with kinetic energies up to several kilo-electron volts~\cite{Opal1971}. For higher kinetic energies $T_{\mathrm{p}}$, relativistic effects as well as close collisions giving rise to the emission of fast secondary electrons (knock-on electrons) are known to modify the f distribution. To account for these effects, we adopt the analytical expression provided in~\cite{Arqueros2009}:
\begin{equation}
\label{eqn:f0}
f_0(T_\mathrm{p},T)=\frac{8\pi ZR_y^2}{m\left(\beta(T_{\mathrm{p}})c\right)^2}\frac{1+C\exp{(-T/T_\mathrm{k})}}{T^2+\overline{T}^2},
\end{equation}
where $c$ is the speed of light, $m$ is the electron mass, $\beta$ is the relativistic factor, $R_y$ is the Rydberg constant, $T$ ranges from 0 to $T_{\mathrm{max}}=(T_{\mathrm{p}}-I_0)/2$ due to the indistinguishability between primary and secondary electrons (with $I_0$ the ionisation potential to create an electron-ion pair in air), the constant $C$ is determined in the same way as in~\cite{Opal1971} so that $\int \mathrm{d}T~f_0(T)$ reproduces the total ionisation cross section, $T_\mathrm{k}=77~$eV is a parameter acting as the boundary between close and distant collisions, and $\overline{T}=11.4~(15.2)~$eV for nitrogen (oxygen)~\cite{Opal1972}. In the energy range of interest, this expression leads to $\left\langle T\right\rangle\simeq 40~$eV, in agreement with the well-known stopping power. It should be noted, however, that the above equation does not include all the relativistic and exchange effects in the energy distribution of ionisation electrons~\cite{Rosado2014}, but these effects are unimportant for a primary kinetic energy of 60~MeV assumed in this work.

The quantity given by equation~\ref{eqn:d2N_dxdT} corresponds to the average number of secondary electrons per length and per kinetic energy unit left just after the passage of one primary high-energy electron in air. The same quantity available at any time $t$ after the passage of one primary high-energy electron is governed by the interactions that these electrons undergo in air. In turn, the evolution in time of the left hand side of eqn.~\ref{eqn:d2N_dxdT} can be fully encompassed in the time dependence of the distribution in kinetic energy $f(T,t)$ of the ionisation electrons (note that, for convenience, we drop hereafter the $T_\mathrm{p}$ dependence). This evolution is determined by a Boltzmann equation accounting for all the interactions of interest at work. Considering the ionisation as quasi-static in space given the low energy of the electrons, it can be shown that the dominant term governing the time evolution of $f$ is the collision one~\cite{AlSamarai2016}:
\begin{eqnarray}
\label{eqn:boltzmann}
\frac{\partial f(T,t)}{\partial t}&=&\frac{\mathcal{N}c}{A} \Bigg[-\sum_{m=\mathrm{N}_2,\mathrm{O}_2}\hspace{-3mm}\rho_m\beta(T)\left(\sigma^{m}_{\mathrm{ion}}(T)+\sigma^{m}_{\mathrm{exc}}(T)\right)f(T,t) \nonumber \\
&-&\hspace{-5mm}\sum_{m=\mathrm{CO}_2,\mathrm{H}_2\mathrm{O}}\hspace{-5mm}\rho_m\beta(T)\sigma^{m}_{\mathrm{exc}}(T)f(T,t)  - \rho_{\mathrm{O}_2}\beta(T)\sigma^{\mathrm{O}_2}_{\mathrm{att}}(T)f(T,t) \nonumber \\
&+&\hspace{-3mm}\sum_{m=\mathrm{N}_2,\mathrm{O}_2} \hspace{-3mm}\rho_m\int_{T}^{T^{\mathrm{max}}}\mathrm{d}T^\prime\beta(T^\prime)\left(\frac{\mathrm{d}\sigma^m_{\mathrm{ion}}}{\mathrm{d}T}(T^\prime, T)+\frac{\mathrm{d}\sigma^m_{\mathrm{ion}}}{\mathrm{d}T}(T^\prime, T^\prime-T-I_0)\right)f(T^\prime,t) \nonumber \\
&+&\hspace{-3mm}\sum_{m=\mathrm{N}_2,\mathrm{O}_2,\mathrm{CO}_2,\mathrm{H}_2\mathrm{O}}\hspace{-5mm} \rho_m\int_{T}^{T^{\mathrm{max}}}\mathrm{d}T^\prime\beta(T^\prime)\frac{\mathrm{d}\sigma^m_{\mathrm{exc}}}{\mathrm{d}T}(T^\prime, T)f(T^\prime,t)\Bigg],
\end{eqnarray}
where $\sigma^m_i$ denotes the cross sections of interest, namely ionisation, excitation of electronic levels (including ro-vibrational excitation) and attachment processes for a molecule $m$. The three first terms in the right hand side stand for the disappearance of electrons with kinetic energy $T$, while the two last terms stand for the appearance of electrons with kinetic energy $T$ due to ionisation and excitation reactions initiated by electrons with higher kinetic energy $T^\prime$. Note that in the case of ionisation, a second electron emerges from the collision with kinetic energy $T^\prime-T-I_0$. 

\begin{figure}[!t]
\centering
\includegraphics[width=12cm]{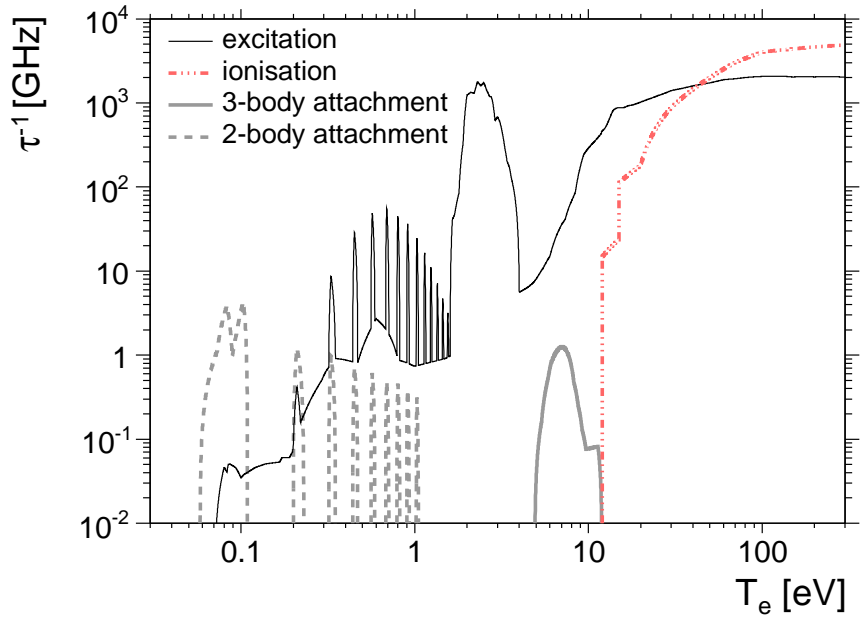}
\caption{\small{Dominant collision rates for low-energy electrons in air, at sea level for a water concentration of 3,000 ppm.}}
\label{fig:rates}
\end{figure}

All cross sections of interest for this study are taken from experimental tabulated data in~\cite{JILA} and are shown in figure~\ref{fig:rates} as a function of the kinetic energy. For completeness, rates are shown  down to kinetic energies much lower than those relevant for the production of UV photons. Going down in energy, the main features of the different rates can be summarised in the following way:
\begin{itemize}
\item For $T\geq 40~$eV, ionisation on N$_2$ and O$_2$ molecules is the dominant process, causing energy losses on a time scale below a picosecond.
\item For 4~eV~$\leq T\leq 40~$eV, excitation on electronic levels of N$_2$ and O$_2$ molecules is the dominant process.  The corresponding energy losses occur on time scales going from picoseconds to a few nanoseconds when going down in energy. To a smaller extent, there is a disappearance of electrons through the three-body attachment process. 
\item For 1.7~eV~$\leq T\leq 4~$eV, resonances for excitation on N$_2$ and O$_2$ molecules through ro-vibrational processes cause energy losses on a time scale of the picosecond.
\item For 0.2~eV~$\leq T\leq 1.7~$eV: 
\begin{itemize}
\item Resonances for excitation on N$_2$ and O$_2$ molecules through ro-vibrational processes quantised in energies are visible through the peaks in continuous line. The corresponding energy losses occur on a time scale of a few tens of picoseconds.
\item Resonances for two-body attachment process on O$_2$ molecules quantised in energies are visible through the peaks in dashed line. The corresponding time scale of disappearance of the electrons is of the order of a nanosecond. This process is subdominant compared to the previous one. 
\item For energies between the quantised ones where ro-vibrational and attachment processes occur, energy losses on the small fractions of CO$_2$ and H$_2$O molecules are important to consider. The corresponding energy losses occur on a time scale of a nanosecond. Note that even if the abundance of Ar in air is larger than for CO$_2$ and H$_2$O molecules, the contribution of Ar molecules to the collision rate is negligible because the corresponding energy losses occur at energies above 1~eV, where collision rates due to N$_2$ and O$_2$ molecules are much more important.
\end{itemize}
\item For $T\leq 0.2~$eV, energy losses on CO$_2$ and H$_2$O molecules degrade electron energies down to 0.1~eV, where the two-body attachment process make them disappearing on a time scale of a few nanoseconds.
\item Although the abundance of Ar in air is larger than for CO$_2$ and H$_2$O molecules, the corresponding energy losses occur at energies above 1~eV, and are thus negligible compared to the energy losses due to collision on N$_2$ and O$_2$ molecules in this energy range. 
\item Other processes such as Bremsstrahlung or recombination on O$_2^+$ ions have much smaller rates, and they are negligible in determining the evolution with time of the kinetic energy distribution of the electrons. Besides, at high energy, electrons can also lose energy in air through K-shell ionisation and nuclear Bremsstrahlung, both processes yielding energetic photons~\cite{Arqueros2009,Rosado2014}. These are however neglected in this study, since we assume that energy is locally deposited in the medium.
\end{itemize}

\begin{figure}[!t]
\centering
\includegraphics[width=12cm]{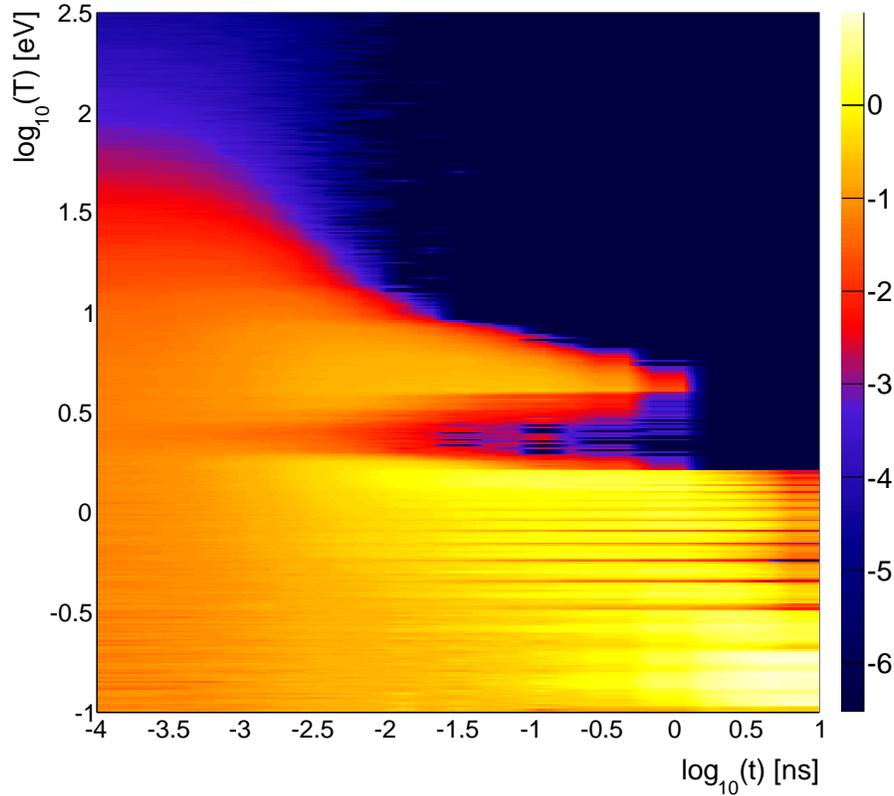}
\caption{\small{The kinetic energy distribution function $f(T,t)$ in log$_{10}$(eV$^{-1}$) units, as a function of time and kinetic energy.}}
\label{fig:f2d}
\end{figure}

To solve eqn.~\ref{eqn:boltzmann}, a Monte-Carlo generator is used to simulate a large number of electrons with initial energy randomly chosen according to $f_0(T_\mathrm{p},T)$. All interactions are then simulated according to the cross sections of interest, until the first test electron and all secondary electrons produced by ionisation get attached to oxygen molecules. This allows us to histogram $f(T,t)$ at any time $t$. The solution for $f$ is shown in figure~\ref{fig:f2d} in a decimal logarithm scale under 800~hPa pressure and 293~K temperature conditions, as a function of $T$ and $t$.
Dips at kinetic energies corresponding to collision rates at quantised energies depicted in figure~\ref{fig:rates} are clearly seen. On a time scale of 100~ps, almost all electron energies are already below the ionisation threshold. After 1~ns, the number of electrons with energies around $\simeq 2$~eV drops off to a large extent, due to excitation reactions. Note that, given the different thresholds of the reactions to produce UV photons, the phase space below 3~eV and above 10~ns will be irrelevant for the purpose of this study.

Having determined the time evolution of the $f$ distribution, the average number of ionisation electrons per unit length and per kinetic energy band can be estimated at any time as
\begin{equation}
\label{eqn:d2N_dxdT_bis}
\frac{\dif^2n_{\mathrm{e,i}}(T,t)}{\dif x~\dif T}=\frac{\rho \mathcal{N}}{A}\sigma_{\mathrm{ion}}f(T,t).
\end{equation}
Of particular interest for the following is to consider the number of electrons per unit time and kinetic energy unit crossing any elementary sphere surface around the primary interaction point. This is obtained by coupling equation~\ref{eqn:d2N_dxdT_bis} to the speed of the particles $\beta(T)c$ and integrating over solid angle:
\begin{equation}
\label{eqn:phi1}
\phi_{\mathrm{e,i}}^1(T,t)=\int\frac{\dif\Omega}{4\pi}\frac{\dif^2n_{\mathrm{e,i}}(T,t)}{\dif x\dif T}\beta(T)c=\frac{\rho\mathcal{N}}{A}c\sigma_{\mathrm{ion}}f(T,t)\beta(T). 
\end{equation}

\section{Fluorescence Yield from Excitation of Nitrogen Molecules}
\label{sec:fy}

In this section, the deposited flux of low-energy ionisation electrons per unit time and kinetic energy unit estimated in the previous section is used to calculate first the UV fluorescence radiation emitted at 337~nm coming from the second positive system (2P) of N$_2$ molecules, and then the total fluorescence radiation emitted in the $[330-400]~$nm wavelength range. For comparison purpose, the energy of the primary electron is taken at an energy corresponding to the electron-energy equivalent assigned to the AIRFLY measurement~\cite{Ave2007} for 120~GeV protons, namely $T_{\mathrm{p}}\simeq 60~$MeV.

\subsection{Fluorescence Yield from $\mathrm{C}~^3\Pi_u$ to $\mathrm{B}~^3\Pi_g$ Radiative Transition in $\mathrm{N}_2$}

Through their collisions, ionisation electrons can excite nitrogen molecules in their ground state $\mathrm{X}~^1\Sigma_g^+$ to upper electronic states. Each electronic state is split into vibrational levels $v$. The set of electronic transitions $v-v^\prime$ between the level $v$ of an upper electronic state and the level $v^\prime$ of a lower state is a particular band system. The band we are interested in here corresponds to the vibrational levels $(v,v^\prime)=(0,0)$ of the N$_2$  $\mathrm{C}~^3\Pi_u\rightarrow\mathrm{B}~^3\Pi_g$ de-excitation, giving  rise to the production of 337~nm wavelength photons. 

The cross section that quantifies the intrinsic likelihood of an ionisation electron exciting a N$_2$ molecule yielding \textit{in fine} to the production of 337~nm wavelength photons is known as the optical cross section, $\sigma_{337}(T)$. Interestingly, direct measurements of this optical cross section are available, within 15\% of uncertainties. The optical cross section as reported in~\cite{Itikawa2005} is used in the following. Hence, the number of 337~nm wavelength photons produced per time and length units is obtained by coupling equation~\ref{eqn:phi1} to this cross section in the following way:
\begin{equation}
\label{eqn:rate_exc}
\nu_{337}(t)=\frac{\rho_{\mathrm{N}_2} \mathcal{N}}{A}\int\dif T ~\phi_{\mathrm{e,i}}^1(T,t)\sigma_{337}(T).
\end{equation}
Therefore, in the absence of other transition processes apart from radiative ones, the fluorescence yield per deposited energy unit $Y^0_{337}$, where the subscript $0$ stands for the absence of quenching effect (see below), is obtained by dividing this rate by $\rho(\dif E/\dif X)_{\mathrm{tot}}$ - the total deposited energy per unit length - and by integrating over time:
\begin{equation}
\label{eqn:yield0}
Y^0_{337}=\left(\frac{\mathcal{N}}{A}\right)^2\frac{c\rho_{\mathrm{N}_2}\sigma_{\mathrm{ion}}}{(\dif E/\dif X)_{\mathrm{tot}}}\int\dif t\int \dif T~\beta(T)f(T,t)\sigma_{337}(T).
\end{equation}
The integration time limit is arbitrary, but has to be greater than the time during which $f$ is non-zero in the kinetic energy range of interest. 

Finally, the effective fluorescence yield $Y_{337}$ is obtained by considering quenching effects, due to non-radiative de-excitations through collisions with other molecules of the medium. This quenching effect can be shown to be accounted for through the introduction of the so-called Stern-Volmer correcting factor in the following way~\cite{Arqueros2009}:
\begin{equation}
\label{eqn:yield}
Y_{337}=Y^0_{337}\left(1+P/P'_{337}\right)^{-1}.
\end{equation}
The parameter $P'_{337}$ is to be interpreted as the characteristic pressure accounting for the collisional quenching. 

The numerical result found for $Y_{337}$ reads
\begin{equation}
\label{eqn:yield_AN}
Y_{337}=(6.05~\pm 1.50)~\mathrm{MeV}^{-1}.
\end{equation}
To get this result, we have used $(\dif E/\dif X)_{\mathrm{tot}}\simeq 2.35$~MeV~g$^{-1}$~cm$^2$~\cite{Berger} and  $P'_{337}\simeq15.9~$hPa~\cite{Ave2007}. Note that the  uncertainty of this result mainly comes from uncertainties in the optical cross section, in the $P'_{337}$ factor, in the cross sections for ionisation, excitation and attachment, and in the energy spectrum of ionisation electrons at their creation time. Within the uncertainties, this result for $Y_{337}$ compares well to previous dedicated works~\cite{Rosado2014} and to experimental results. For comparison purpose, we note that the world average value obtained in~\cite{Rosado2014} from various experiments accounting for numerous effects to compare the results under the same pressure and temperature conditions is $\left\langle Y_{337}\right\rangle=(7.04\pm0.24)~\mathrm{MeV}^{-1}$.



\subsection{Fluorescence Yield in the $[330-400]~\mathrm{nm}$ Wavelength Band}

The 337 nm radiative transition is the dominant process contributing the air-fluorescence, but other radiative transitions enter into play in the wavelength band of interest. Considering the efficiency of the fluorescence telescopes as roughly rectangle functions between 330~nm and 400~nm, the fluorescence yield for this wavelength interval can be found by dividing $Y_{337}$ with the ratio of the fluorescence emission. The fluorescence emission ratio is the fluorescence emission $I_{337}$ at wavelength $\lambda=337~$nm, to the fluorescence emission for the desired wavelength interval $\Delta\lambda=[330-400]~$nm. The fluorescence emission factors can be obtained from the measured intensities reported in~\cite{Ave2007}, leading to $Y_{[330-400]}\simeq17.6~$~MeV$^{-1}$. This is the relevant figure to which the contribution of molecular Bremsstrahlung radiation to the fluorescence yield estimated in next section in this wavelength interval has to be compared.

\section{Fluorescence Yield from Molecular Bremsstrahlung Radiation}
\label{sec:fy_mbr}

Through their collisions, ionisation electrons can produce their own emission through the molecular Bremsstrahlung process. The production of photons with energies $h\nu$, with $h$ the Planck constant and $\nu$ the frequency, corresponds to transitions between unquantised energy states of the electrons. Screening effects prevent quantitative calculations of the Bremsstrahlung cross section without approximations which depend on the  kinetic energy of the electron and on the emitted energy of the photon. For $T$ up to a few tens of electron volts, screening effects are important and the differential cross section $\dif\sigma_{\mathrm{MBR}}/\dif\nu$ is well described by the following expression valid for photon frequencies greater than the rate of successive collisions of electrons obtained in~\cite{Kasyanov1961}:
\begin{equation}
\label{eqn:sigma_ff}
\frac{\dif\sigma_{\mathrm{MBR}}(T,\nu)}{\dif\nu}=\frac{4\alpha^3}{3\pi R_y\nu}\left(1-\frac{h\nu}{2T}\right)\left(1-\frac{h\nu}{T}\right)^{1/2}T\sigma_{\mathrm{m}}(T),
\end{equation}
where $\alpha$ is the fine-structure constant, $R_y$ the Rydberg constant and $\sigma_{\mathrm{m}}$ the electron transfer cross section taken from tabulated data in~\cite{JILA}. For $T\leq100~$keV and $h\nu/T\ll 1$, the photon emission resulting from this process is isotropic. 

The corresponding contribution to the fluorescence yield can be roughly\footnote{We neglect here the transition radiations associated to the appearance and disappearance of the electrons.} estimated in the same way as in the previous section, by plugging in the Bremsstrahlung cross section instead of the optical one and by integrating over the relevant frequency range. This leads to the following expression:
\begin{equation}
\label{eqn:yield_mbr}
Y_{[330-400]}^{\mathrm{MBR}}=\left(\frac{\mathcal{N}}{A}\right)^2\frac{c\rho\sigma_{\mathrm{ion}}}{(\dif E/\dif X)_{\mathrm{tot}}}\int\dif t\int \dif T\int \dif\nu~\beta(T)f(T,t)\frac{\dif\sigma_{\mathrm{MBR}}(T,\nu)}{\dif\nu}.
\end{equation}
Note that, to restrict to a compact expression for convenience, the differences in the momentum transfer cross section for nitrogen and oxygen target molecules have been omitted in this expression. The actual calculation splits this equation into two terms, weighted by the densities of nitrogen and oxygen. Once integrations are carried out, the obtained result is
\begin{equation}
\label{eqn:yield_mbr_AN}
Y_{[330-400]}^{\mathrm{MBR}}=1.0\times10^{-1}~\mathrm{MeV}^{-1}.
\end{equation}
Hence, compared to $Y_{[330-400]}$, it turns out that molecular Bremsstrahlung radiation contributes $\simeq 0.5\%$ to the fluorescence yield.  This is a quite small contribution, which nevertheless is not negligible in regards of the total budget of current systematic uncertainties to consider for the absolute energy scale of the fluorescence detector~\cite{Verzi2013}.

\section{Conclusions}
\label{sec:conclusions}

In this study, the contribution of molecular Bremsstrahlung radiation to the fluorescence yield has been presented, based on a general approach that follows from previous works dedicated to the emission in the gigahertz frequency range. Applied to the de-excitation processes in N$_2$ molecules leading to the fluorescence emission in the UV range, the approach is shown to reproduce the main features of the fluorescence emission in air. Applied to the molecular Bremsstrahlung radiation of the low-energy ionisation electrons, the approach provides the estimate that this process contributes $\simeq 0.5\%$ to the fluorescence yield in the $[330-400]~$nm wavelength range under 800~hPa pressure and 293~K temperature conditions. This means that out of $\simeq 175~$photons with wavelength between 330 and 400~nm detected by the fluorescence detectors, one of them has been produced by molecular Bremsstrahlung radiation. Note that in extensive air shower, this contribution slightly increases with decreasing altitude because of the fluorescence quenching.  

Hence, not surprisingly since the emission induced by de-excitation processes is known to reproduce the main features of the fluorescence, the contribution of molecular Bremsstrahlung radiation to the fluorescence yield appears as a small one. It is however to be noted that this contribution is of the same order of magnitude as other sources of systematic uncertainties to consider for the absolute energy scale of fluorescence detectors.

\section*{Acknowledgements}

We acknowledge the support of the French Agence Nationale de la Recherche (ANR) under reference ANR-12-BS05-0005-01, and of the Spanish Ministry of Economy and Competitiveness (MINECO) under reference FPA2012-39489-C04-02.
We thank Roger Clay for his careful review of this paper, and Valerio Verzi for his suggestions and encouragements.


\end{document}